\documentclass[%
 reprint,
 superscriptaddress,
 amsmath,amssymb,
 aps,pre,
floatfix
]{revtex4-1}

\usepackage{graphicx}
\usepackage{dcolumn}
\usepackage{bm}
\usepackage{color}
\usepackage[caption=false]{subfig}
\usepackage{footmisc}

\begin{document}
\preprint{APS/123-QED}

\title{Two classes of events in sheared particulate matter}

\author{Peter K. Morse}
\affiliation{Department of Physics, Syracuse University, Syracuse, New York 13244, USA }
\author{Sven~Wijtmans}
\affiliation{Department of Physics, Syracuse University, Syracuse, New York 13244, USA }
\author{Merlijn~van Deen}
	\affiliation{Huygens-Kamerlingh Onnes Lab, Universiteit Leiden, P.O.~Box~9504, NL-2300 RA Leiden, The Netherlands}
\author{Martin~van Hecke}
	\affiliation{Huygens-Kamerlingh Onnes Lab, Universiteit Leiden, P.O.~Box~9504, NL-2300 RA Leiden, The Netherlands}
	\affiliation{AMOLF, Science Park 104, 1098 XG Amsterdam, The Netherlands}
\author{M. Lisa~Manning}
\affiliation{Department of Physics, Syracuse University, Syracuse, New York 13244, USA }
\date{\today}

\begin{abstract}
Under shear, a system of particles changes its contact network and becomes unstable as it transitions between mechanically stable states. For hard spheres at zero pressure, contact breaking events necessarily generate an instability, but this is not the case at finite pressure, where we identify two types of contact changes: network events that do not correspond to instabilities and rearrangement events that do. The relative fraction of such events is constant as a function of system size, pressure and interaction potential, consistent with our observation that both nonlinearities obey the same finite-size scaling. Thus, the zero-pressure limit of the nonlinear response is highly singular.
\end{abstract}


\maketitle

Under shear, systems of interacting particles undergo a sequence of 
transitions between mechanically stable states. For 
supercooled liquids and glassy solids, localized excitations that allow the material to flow and fail under an applied strain correspond to low-energy saddle points of this landscape~\cite{brito_geometric_2009, manning_vibrational_2011, dasgupta_universality_2012}.
While this scenario is well-established for materials with attractive interactions or at significant finite pressures, it is not clear whether this scenario holds for pressures close to zero.

The physics of weakly compressed systems of repulsive soft particles is governed by the jamming transition.
Linear response properties, such as the shear and bulk moduli,  
robustly follow finite size scaling as a function of the number of particles $N$ and pressure $p$~\cite{goodrich_finite-size_2012, dagois-bohy_soft-sphere_2012, corwin_viewpoint:_2012, goodrich_jamming_2014, goodrich_scaling_2016}.  These and other results for the linear response suggest that jamming is a bona fide phase transition. The number of contacts between particles plays a crucial role:
at zero pressure, systems of repulsive spheres are isostatic, so that the number of degrees of freedom equals the number of constraints. The mechanical response of such systems is highly anomalous, and
the removal of any contact will destabilize the system~\cite{maxwell_calculation_1864, calladine_buckminster_1978, lubensky_phonons_2015}.
At finite pressure, the excess number of contacts above isostaticity $N \Delta Z$ also is governed by finite size scaling, revealing a small pressure regime with only one excess contact ~\cite{goodrich_finite-size_2012, dagois-bohy_soft-sphere_2012, goodrich_jamming_2014, goodrich_scaling_2016}.

Here we investigate jammed systems under large shear deformations,
focusing on contact changing events \cite{van_deen_contact_2014, van_deen_contact_2016}. Near the critical point, breaking a single contact might have a dramatic effect. In particular, at zero pressure, any contact breaking corresponds to a loss of stability, while at finite pressures, where the excess number of contacts is positive, it could be possible to break a contact without initiating an instability. It is however not yet clear how often this happens, or even whether it is possible to distinguish between true instabilities that correspond to a saddle point, and innocuous contact changes not associated with a saddle point.

We find that at finite pressures, we can sharply distinguish two classes of contact changes: {\em network events}, which do not correspond to instabilities, and {\em rearrangements}, which correspond to saddles in the energy landscape. Rearrangements are associated with finite jumps in both the shear stress and irreversibiliy, while network events are not.
We study the frequency of network events and rearrangements as a function of $N$ and $p$  and find that the fraction of all contact changes that correspond to instabilities is remarkably constant as a function of $N$, contact potential, and $p$. This highlights that while all contact breaking events are instabilities at isostaticity, at all pressures above isostaticity the two are abruptly no longer one-to-one.

Furthermore, we find that near isostaticity the nonlinear properties, such as stress jumps, degree of irreversibility associated with rearrangements, and the mean strain steps between subsequent network events and subsequent rearrangements, exhibit the same finite-size scaling that has previously been reported for linear response~\cite{goodrich_scaling_2016}. Specifically, the scaling exponents extracted from simulations appear to be integer,
dependent on the interaction potential (Hertzian or Hookean), and can be rationalized using simple arguments. This surprising correspondence between the linear and nonlinear scaling is similar to what is found in Random Field Ising Models~\cite{kent-dobias_cluster_2018}.

Finally, Wyart has recently argued that if finite contact potentials at small pressure introduce small perturbations to the hard-sphere system, then one expects the minimum force $f_{min}$ across a Hertzian contact at the onset of instability to scale as $p^3$~\footnote{Unpublished work by Matthieu Wyart}.  With the correct finite-size scaling, we are able to resolve the statistics of the minimum force at very low pressures and find that it instead scales as $f_{min} \sim p^2$.

Taken together, our results highlight that the zero pressure limit, which corresponds to the physics of hard spheres, is highly singular and cannot by itself describe the nonlinear finite-pressure behavior of particulate matter in low dimensions. In other words, the phenomenology of jamming at small but finite pressures is different from that at zero pressure.

{\em Model.---} We simulate systems of 2D bidisperse disks in a 50-50 mixture with size ratio 1:1.4 in a square box under athermal quasistatic shear \cite{maloney_amorphous_2006}. We perform an infinite temperature quench to obtain an initial packing and use pairwise particle contact potentials to define the energy: $ U = \sum_{ij}\Theta(\varepsilon_{ij})\varepsilon_{ij}^{\xi}$
where $\varepsilon_{ij}$ is the overlap between particles i and j, $\Theta$ is the heaviside function, and $\xi$ represents the power of the potential. The stress $\sigma$ is calculated via the Born-Huang approximation \cite{born_dynamical_1954}.
We study Hertzian ($\xi = 5/2$) and Hookean ($\xi = 2$) disks 
in 2D and Hertzian spheres in 3D, to understand whether the two different types of potentials, which generate distinct interparticle stiffnesses as a function of pressure, affect our results.

\begin{figure}
\includegraphics[width=0.95\columnwidth]{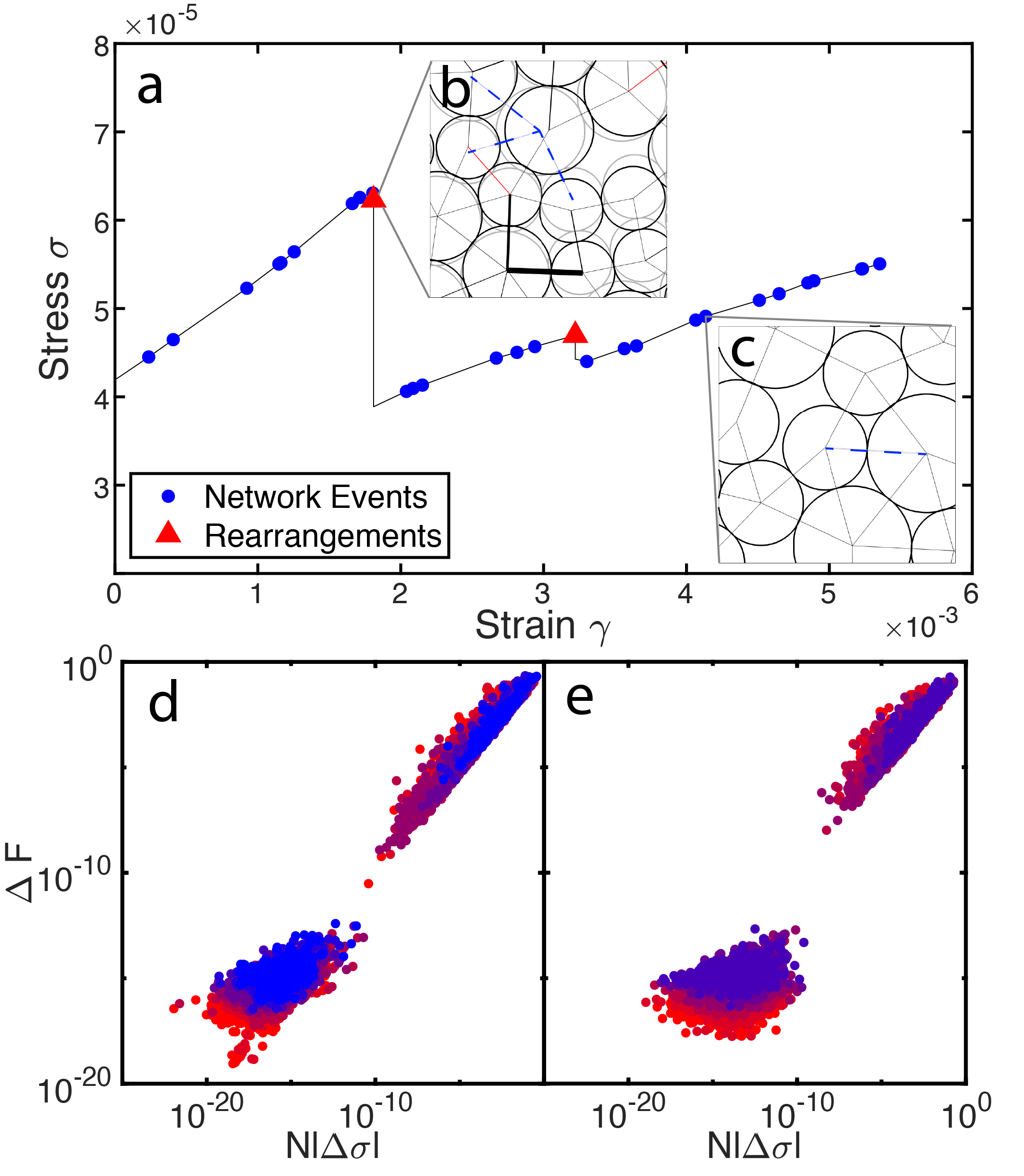}
\caption{\textbf{a)} Typical stress strain curve for 2D Hertzian spheres with $N=128$ and $p=10^{-2}$. Network events are labelled with blue dots and rearrangement events are labeled with red triangles. Examples of particle positions and forces during \textbf{b)} a rearrangement and \textbf{c)} a network event, where grey disks represent positions before and solid disks represent positions after the event. Solid red lines represent new contacts and dashed blue lines represent broken contacts. A metric of reversibility versus the stress drop for \textbf{d)} 2D Hertzian and \textbf{e)} 2D Hookean disks distinguishes network events (lower left) from rearrangements (upper right) for all system sizes and pressures, where the colors represent N = 32(red), 64, 128, 256, 512, and 1024(blue) with an even gradient, and pressures are not distinguished.}
\label{fig:stressDropsDef}
\end{figure}

To apply a shear-strain $\gamma$, we utilize standard athermal quasi-static shear with Lees-Edwards boundary conditions where the energy is minimized after each strain step via the FIRE algorithm \cite{bitzek_structural_2006} until the maximum unbalanced force on any particle is less than $10^{-18}$. 
As we apply shear, the system undergoes contact changes, and we use a bisection algorithm to bracket contact changes \cite{van_deen_contact_2014} with a resolution of $\Delta \gamma < 10^{-13}$ \footnote{There is a finite density of particles with less than $D+1$ contacts, which are termed rattlers.
In our statistics,  we only consider particles which are not rattlers either before and after an event.}.
As we frequently need to distinguish small numbers from those that are zero to numerical precision, we employ quad precision calculations and perform the minimizations on GPUs \footnote{pyCudaPack available upon request at https://github.com/simonsglass/pycudapacking/} to access a broad range of parameters with high numerical precision.

\begin{figure}
\includegraphics[width=0.95\columnwidth]{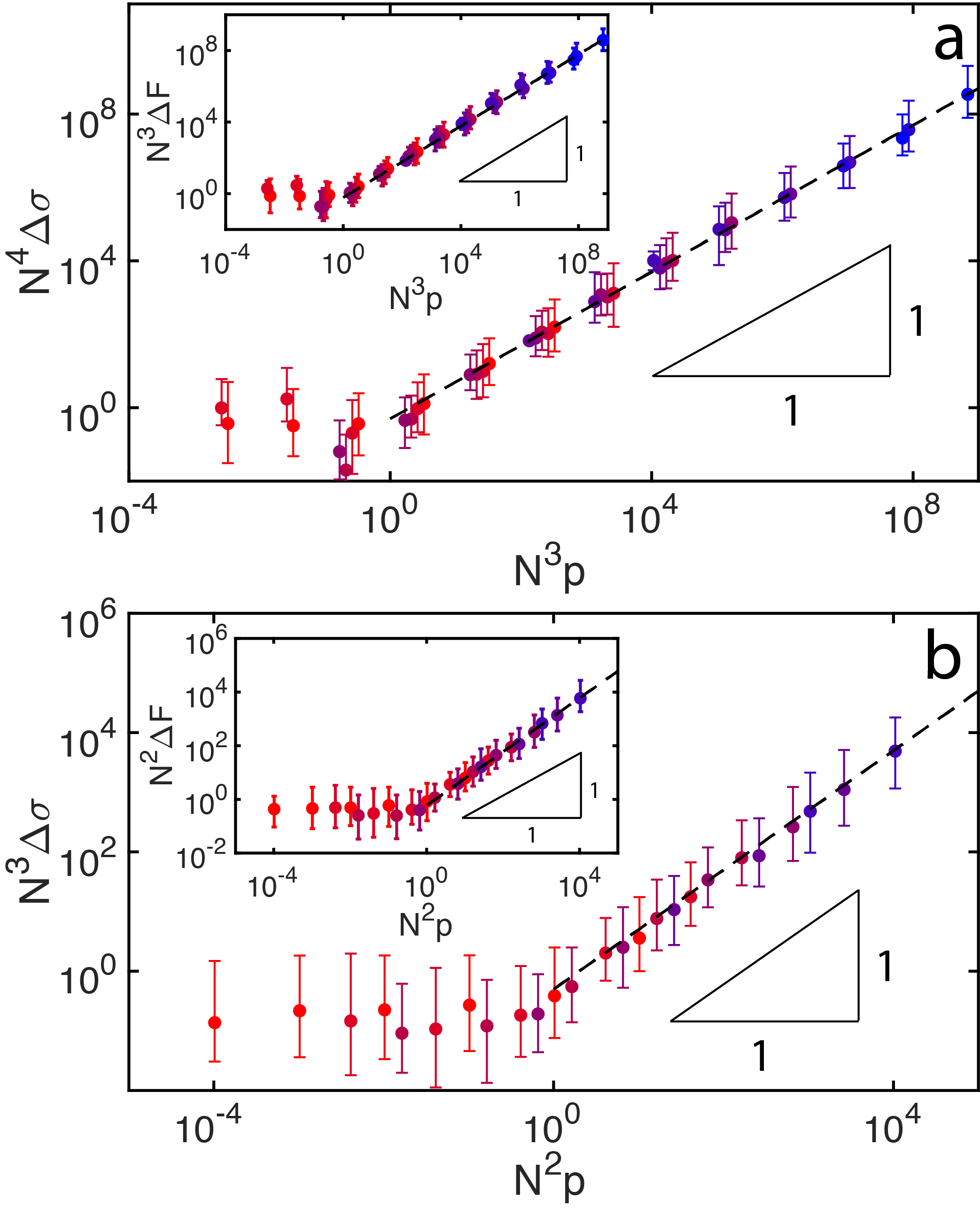}
\caption{(color online) The average stress drop associated with rearrangements for each system size and pressure collapses onto a master curve when we plot it versus $N^\beta p$ for (a) 2D Hertzian disks ($\beta = 3$) and (b) 2D Hookean disks ($\beta = 2$). \textbf{(insets)} The metric of reversibility associated with rearrangements also collapses onto a master curve when plotted versus $N^\beta p$. Dashed lines have slope unity and are shown to guide to the eye. Error bars represent the middle 60\% of each distribution, and the colors are N = 32(red), 64, 128, 256, 512, and 1024(blue) with an even gradient.}
\label{fig:allCrosshair}
\end{figure}

{\em Two classes of events.---}
We study contact changes in ensembles of 50 different initializations at pressures ranging from $10^{-7}$ to $10^{-2}$ and systems sizes from 32 to 4096.
For each case, we apply shear until there have been either 30 contact changes in 2D or 60 contact changes in 3D, which allows us to stay within the pre-yielding regime. Our first main result is that all contact changes are of 
two mutually distinct types (Fig.~ \ref{fig:stressDropsDef}). First, we observe events where the stresses exhibits a finite jump and the particles undergo discontinuous motion (Fig.~1b). We refer to these are {\em rearrangements}.  Second, we observe events where individual contacts are broken or created but the stress remains smooth (Fig.~1c). We refer to these as {\em network events}. 

To investigate the nature and statistics of these events, we focus on infinitesimal strain loops associated with the smallest bisection interval around a contact change event. First, the stress drop $\Delta \sigma$  associated with a contact change is defined using a linear extrapolation of the shear modulus $G$: $\Delta \sigma = (\gamma_{+} - \gamma_{-})G_{-} - (\sigma_{+} - \sigma_{-})$, where
the indices $-$ and $+$ represents the last point before the contact change and the first point after the contact change.  Second, to investigate the (ir)reversibility of each contact change,
we study the change in the force contact network, defined using the quadrature sum of the differences in inter-particle forces before and after an infinitesimal strain loop: $\Delta F(\gamma_n)  = \big(\Sigma_{k} |\textbf{f}^{(k)}_{\rightarrow} - \textbf{f}^{(k)}_{\leftarrow}|^2\big)^{1/2}$, where $k$ is a sum over all contact forces and the right (left) arrow indicates the force at the beginning (end) of the loop across the contact change event.

Scatter plots of $\Delta F$ vs $|\Delta \sigma|$ reveal that network events and rearrangements are
clearly distinguished (Fig.~\ref{fig:stressDropsDef}d,e): Network events exhibit no stress drops to within numerical precision $\Delta \sigma < 10^{-14}$, while stress drops associated with rearrangements are finite
\footnote{While we cannot rule out the existence
of saddle points that are not associated with contact changes a priori,
we have found no evidence of such events, and our data places an upper bound on the stress drops that are associated with such events of $N\Delta\sigma < 10^{-10}$.}.

Our data unambiguously shows that network events are perfectly reversible, while rearrangements are irreversible. Hence the two classes of events can be separated by stress drop and (ir)reversibility. Analogous results are true for 3D Hertzian spheres (Fig. S2).

We note that while previous work suggested that $|\Delta \sigma|$ exhibits power-law scaling~\cite{salerno_avalanches_2012}, implying
that arbitrarily small stress drops are possible, more recent work 
~\cite{franz_mean-field_2017} suggests a finite cutoff to the smallest possible stress drops. Our simulations are consistent with the latter scenario, and the existence of a finite minimum magnitude of stress drops associated with saddles (Supplemental Fig S1) makes it possible to distinguish network events from rearrangements.



Together, our results indicate a major difference between the physics at {\em finite} albeit small pressures, and the hard sphere scenario at strictly zero pressure. For hard spheres at zero pressure, contact breaking events will necessarily correspond to a saddle point or instability \cite{wyart_marginal_2012}, which is clearly not the case at any finite pressure.


{\em Finite size scaling.---}
Previously, Goodrich et al. showed the finite-size scaling of linear elastic response -- the shear and bulk moduli -- collapses as a function of $N^\beta p$ \cite{goodrich_finite-size_2012, dagois-bohy_soft-sphere_2012}, implying that jamming has the finite size scaling properties of a real phase transition \cite{widom_equation_1965, fisher_scaling_1972, goodrich_jamming_2014, goodrich_scaling_2016, sastry_critically_2016}. 

Therefore, we use our data to study the statistics of stress drops $\Delta \sigma$ and irreversibility in the force network $\Delta F$ across saddle points at very small pressures, and test the scaling ansatz developed initially for linear response. 


We first extend the previous scaling ansatz for $\Delta Z$ to generic contact potentials~\cite{ohern_random_2002, ohern_jamming_2003}:
\begin{equation}
\Delta Z = \frac{1}{N}W(N^{2(\xi-1)}p) = \frac{1}{N}W(N^{\beta}p)
\end{equation}
\noindent
where $\beta = 3$ for Hertzian systems and $\beta = 2$ for Hookean systems. Systems with $N^\beta p < 1$ have on average only one contact above isostaticity, and exhibit a different behavior than those with $N^\beta p > 1$. Figures 2a and 2b illustrate that the scaling collapse of the nonlinear response -- including both stress drops $\Delta \sigma$ and force network irreversibility $\Delta F$ across saddle points -- is consistent with the previously reported scaling of linear response near jamming~\cite{goodrich_finite-size_2012}. 

\begin{figure}
\includegraphics[width=0.95\columnwidth]{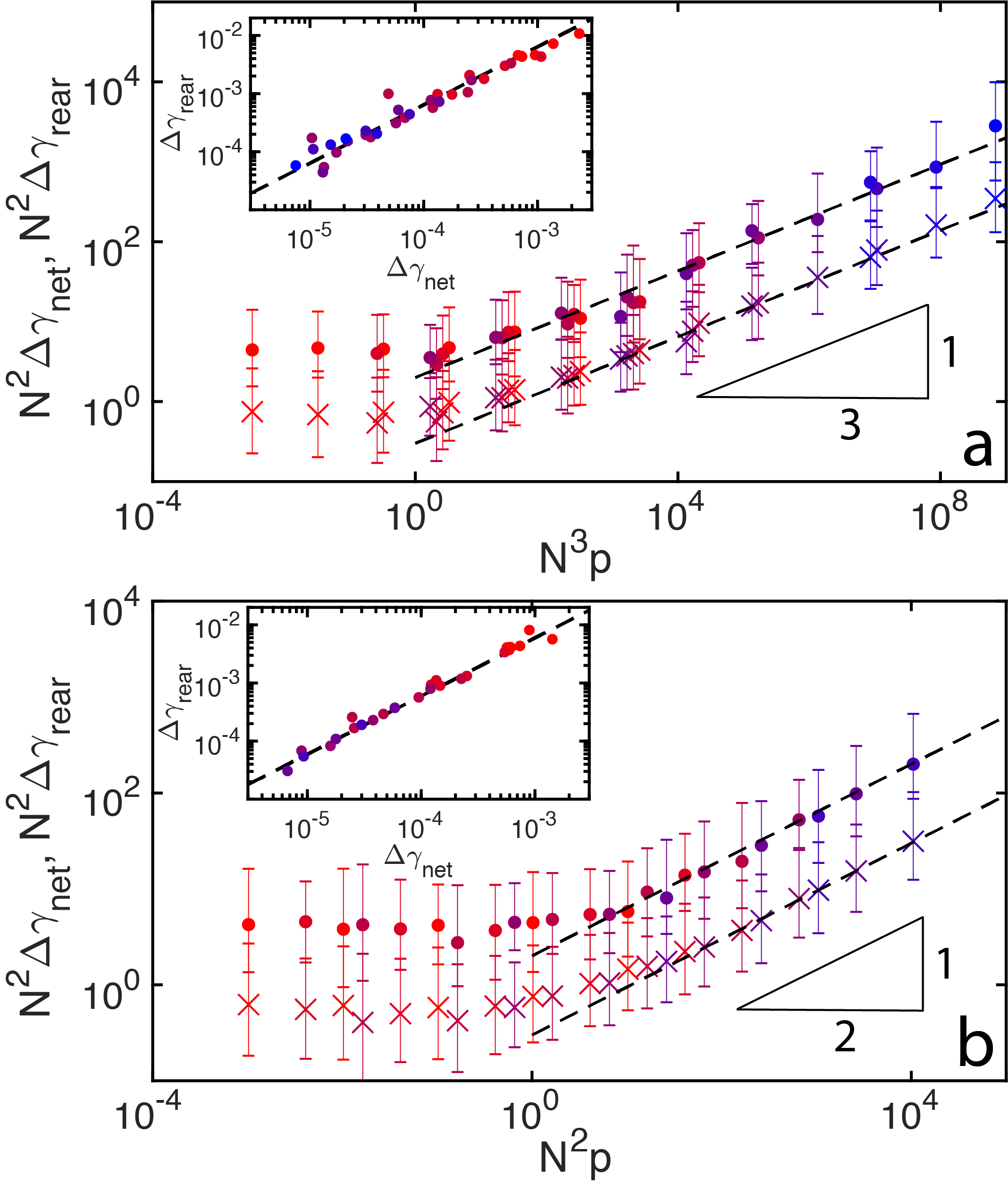}
\caption{Strain steps between rearrangements (circles) and network events (crosses) for all system sizes and pressures of \textbf{a)} 2D Hertzian disks and \textbf{b)} 2D Hookean disks. Colors indicate system size with colors are N = 32(red), 64, 128, 256, 512, 1024, 2048, and 4096(blue) with an even gradient. Error bars represent the middle 60\% of each distribution, and the point represents the geometric mean. Dashed lines with slope 1 are added as a guide to the eye. \textbf{Insets)} A simple regression confirms that these two curves have the same slope, which implies a constant fraction of rearrangements across system size and pressure. }
\label{fig:propReHertz}
\end{figure}

In the region where $N^\beta p > 1$, we find that the scaling of stress drop and the irreversibility metric before rearrangements scale with $N$ and $p$ as $\langle\Delta \sigma \rangle \sim \frac{p}{N}$ and $\langle \Delta F \rangle \sim p$ respectively. The linear scaling of both with pressure implies that the only force scale in the system is the one set by the pressure. And while the typical size of a stress drop goes to zero with increasing system size, the size of the irreversibility metric remains constant, implying that rearrangements involve a characteristic number of particles in the pre-yielding regime.

\begin{figure}
\includegraphics[width=0.95\columnwidth]{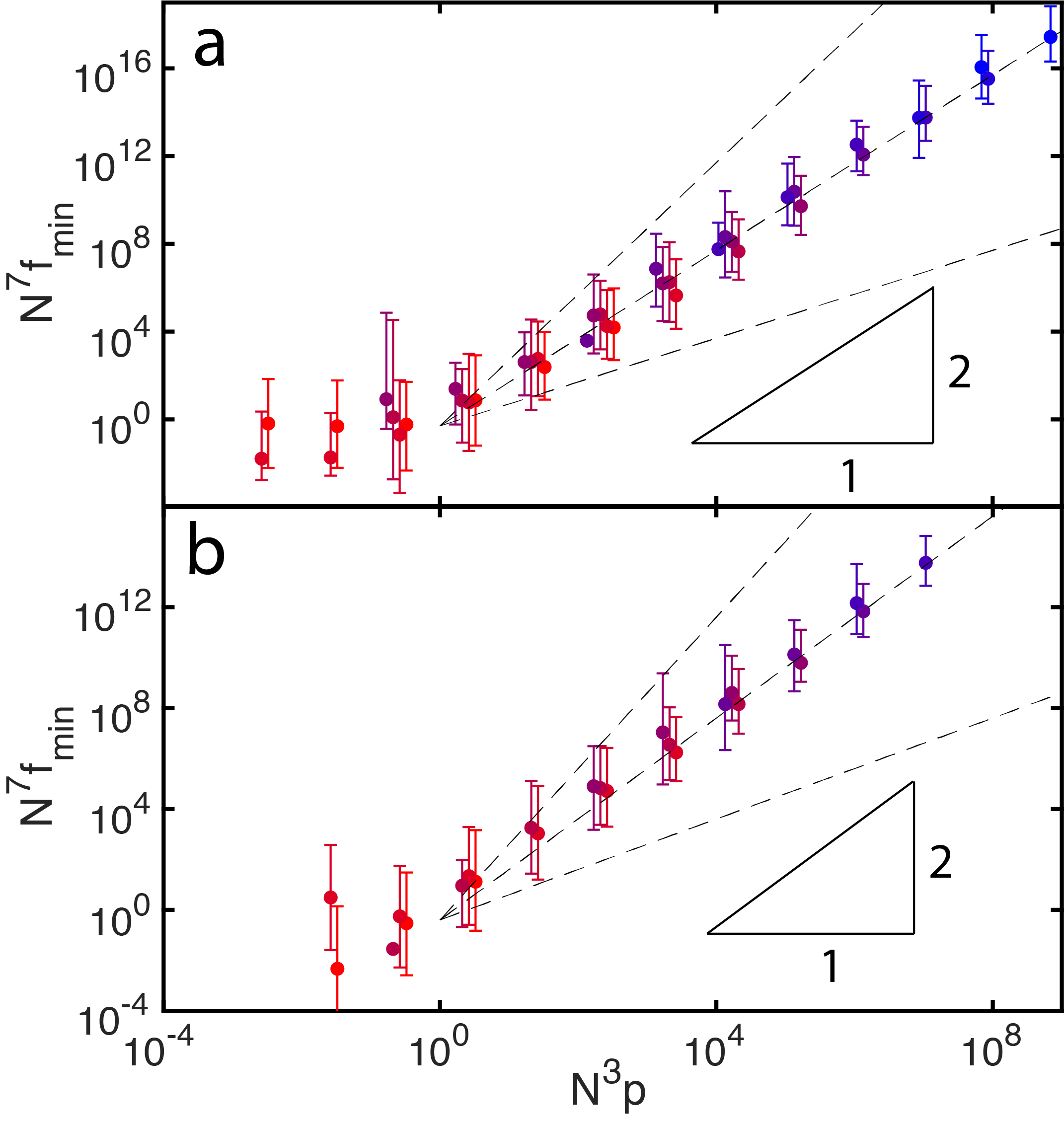}
\caption{The minimum force before a rearrangement in \textbf{a)} 2D Hertzian and \textbf{b)} 3D Hertzian systems. Dashed lines represent a slope of 1, 2, and 3. Our data is consistent with $f_{min}\sim p^2/N$. Error bars represent the middle 60\% of each distribution, and the data point represents the geometric mean. The colors vary with system size as N = 32(red), 64, 128, 256, 512, 1024, 2048, and 4096 (blue) with an even gradient.}
\label{fig:finSizeHertz}
\end{figure}

{\em Characteristics Strain Scales.---} In order to measure the relative frequency of network events and rearrangements, we measure the average strain between network events $\Delta \gamma_{net}$ and the average strain between rearrangements $\Delta \gamma_{rear}$ and analyze the scaling of both as a function of $N^\beta p$ (Fig.~\ref{fig:propReHertz}(a,b)). We again find excellent scaling collapse. Moreover, the fraction of contact changes corresponding to rearrangements remains constant at all pressures
and system sizes. 

Maloney and Lemaitre \cite{maloney_amorphous_2006} predicted the exponent for strain between rearrangements in steady-state flow by arguing that all accumulated energy must be dissipated, so $\Delta\gamma_{rear} \sim \frac{\Delta\sigma}{G}$. Given the numerical observation that the shear modulus scales with the pressure and interaction potential as $G \sim p^{(\xi-3/2)/(\xi-1)}$ \cite{ohern_jamming_2003}, we obtain $\Delta\gamma_{rear} \sim \frac{p^{1/2(\xi-1)}}{N} = \frac{p^{1/\beta}}{N}$.

Surprisingly, Figure \ref{fig:propReHertz} shows this applies even in the pre-yielding regime before steady-state flow, and not only to the distance between rearrangements $\Delta\gamma_{rear}$, but also to the distance between network events $\Delta\gamma_{net}$. This is in agreement with the results of previous work by van Deen \cite{van_deen_contact_2016}, which we extend in the supplement (Supplemental section B.)

A simple regression analysis between $\Delta\gamma_{rear}$ and $\Delta\gamma_{net}$ for Hertzian 2D disks yields $\Delta\gamma_{rear}/\Delta\gamma_{net} = 5.88 \pm 0.35$ (Figure 3a inset). We can convert this to a relative fraction of rearrangements: $\eta = 0.135 \pm 0.015$ for Hertzian 2D disks, $\eta = 0.145 \pm 0.010$ for Hookean 2D disks, and $\eta = 0.143 \pm 0.015$ for Hertzian 3D spheres. These fractions are statistically indistinguishable, implying that the fraction of rearrangements does not depend on the system size, pressure, or interaction potential, and does not seem to vary between 2D and 3D.

Wyart recently predicted the low-pressure scaling of the minimum force across rearranging contacts, by assuming that small overlaps at low pressures generate a small perturbation to the energy landscape of hard spheres at zero pressure~\cite{Note1}.

As the forces across contacts in Hookean packings do not depend on the overlap, we expect $f_{min}$ not to correlate with pressure in this case, which is consistent with our data shown in Supplemental Figure S4.

For Hertzian spheres in 2D and 3D
we measure the minimum force before an event, and find that

\begin{equation}
\langle f_{min} \rangle \sim \frac{p^2}{N},
\end{equation}

as shown in Fig~\ref{fig:finSizeHertz}.
Our data are not consistent with the prediction from a small-overlap expansion around hard spheres, which for a Hertzian interaction potential is $f_{min} \sim p^3$.  These results also suggests the hard sphere limit is singular.

{\em Discussion.---}We have identified two types of contact changes that exist in finite pressure systems: {\em network events} and {\em rearrangements}.

First, the statistics of each are governed by finite size scaling relations with integer exponents,
corresponding to those of linear response quantities. It is far from obvious that the linear and nonlinear response should obey the same scaling ansatz, although
similar relationships have been observed in other systems~\cite{kent-dobias_cluster_2018}. In jammed particulate matter, it could be that
the low-frequency localized excitations that contribute
to the linear response in low dimensions also govern the
transition states, as postulated previously~\cite{karmakar_statistical_2010, manning_vibrational_2011, gartner_nonlinear_2016}.
Second,  the relative proportion of rearrangements
is small (about 14 \%), and independent of interaction potential, system size, and pressure. 
Third, the
minimum force at the onset of instability scales differently 
than predicted  based on an small-overlap expansion around the hard-sphere zero-pressure
state~\cite{Note1}.

Together, these results highlight that the zero-pressure isostatic limit of purely repulsive particles is highly singular.  At zero pressure, every contact breaking event should correspond to an instability, but Fig. 3 demonstrates that at any finite pressure and system size the fraction of contact changes that are instabilities is a fixed constant far from unity. This may come from the fact that in soft spheres, the limit of zero pressure always corresponds to one excess contact, whereas in hard spheres, there are zero excess contacts.

Looking forward, it will be interesting to study whether the vibrational spectrum of soft spheres at low pressure are different in important ways from those of hard spheres at zero pressure.  The eigenvalue spectrum seems to be well-described by an expansion around zero pressure -- Wyart et al \cite{wyart_geometric_2005} and Goodrich et al~\cite{goodrich_finite-size_2012} demonstrate that isostatic systems at zero pressure must possess soft, system-spanning modes associated with breaking a single contact, and that these extended modes give rise to a boson peak in the density of states that shifts to higher frequencies as the pressure increases away from zero.  

The eigenvector spectrum remains more mysterious. A recent manuscript suggests that extended modes found at zero pressure also destabilize the system at small but finite pressures, and that instabilities continuously become more localized as the pressure increases~\cite{shimada_spatial_2018}. However, other work hints that there might be different types of instabilities in 2D and 3D systems.  Charbonneau and collaborators~\cite{charbonneau_jamming_2015} studied the statistics of small forces in jammed hard-sphere packings as a function of dimension, and found that they scale as a power law predicted by mean-field~\cite{wyart_marginal_2012, lerner_low-energy_2013, degiuli_force_2014} and infinite-dimensional~\cite{charbonneau_fractal_2014} theories, provided that localized {\em buckler} excitations were removed. The existence of these localized excitations, which are much more prevalent in 2D and 3D, suggest that mean-field extrapolations from zero pressure may not fully explain the mechanical response. It would be interesting to study whether the finite-size scaling of $f_{min}$ we observe, which is not consistent with zero-pressure mean field predictions, might instead be consistent with a different, localized class of excitations that govern instabilities in low dimensions.

\bibliography{main}

\end{document}


\title{Two classes of events in sheared particulate matter: Supplemental Material}
\author{Peter K. Morse}
\affiliation{Department of Physics, Syracuse University, Syracuse, New York 13244, USA }
\author{Sven~Wijtmans}
\affiliation{Department of Physics, Syracuse University, Syracuse, New York 13244, USA }
\author{Merlijn~van Deen}
	\affiliation{Huygens-Kamerlingh Onnes Lab, Universiteit Leiden, P.O.~Box~9504, NL-2300 RA Leiden, The Netherlands}
\author{Martin~van Hecke}
	\affiliation{Huygens-Kamerlingh Onnes Lab, Universiteit Leiden, P.O.~Box~9504, NL-2300 RA Leiden, The Netherlands}
	\affiliation{AMOLF, Science Park 104, 1098 XG Amsterdam, The Netherlands}
\author{M.Lisa~Manning}
\affiliation{Department of Physics, Syracuse University, Syracuse, New York 13244, USA }

\maketitle

\setcounter{figure}{0}
\subsection{Additional methods}

{\em Energy minimization details---}
The gradient of the energy is given by

\begin{equation}
\frac{\partial{U}}{\partial{r_{i}^{\alpha}}} = -F_i^\alpha = -\sum_{k\in\partial{i}}\varepsilon_{ik} ^{\xi-1}\frac{n_{ik}^{\alpha}}{\sigma_{ik}},
\end{equation}
\noindent
where $r_i^\alpha$ is the displacement associated with moving particle $i$ in the $\alpha$ direction. To find the linear stability of systems we calculate the eigenvalues of the dynamical matrix $H^{\alpha\beta}_{ij} = \frac{\partial^2 U}{\partial{r_{i}^{\alpha}} \partial{r_{j}^{\beta}}}$

\begin{eqnarray*}
H^{\alpha\beta}_{ij} &= \delta_{ij} \sum_{k\in\partial{i}}\bigg[(\xi-1)\frac{\varepsilon_{ik}^{\xi-2}}{\sigma_{ik}^2}n^{\alpha}_{ik}n^{\beta}_{ik} + \frac{\varepsilon_{ik}^{\xi-1}}{\sigma_{ik}\rho_{ik}}(n^{\alpha}_{ik}n^{\beta}_{ik}-\delta^{\alpha\beta})\bigg] \\
&\quad - \delta_{\langle ij \rangle}\bigg[(\xi-1)\frac{\varepsilon_{ij}^{\xi-2}}{\sigma_{ij}^2}n^{\alpha}_{ij}n^{\beta}_{ij} + \frac{\varepsilon_{ij}^{\xi-1}}{\sigma_{ij}\rho_{ij}}(n^{\alpha}_{ij}n^{\beta}_{ij}-\delta^{\alpha\beta})\bigg],
\end{eqnarray*}
\noindent
where $n_{ij}^\alpha$ is the $\alpha$ component of the normal vector between particles $i$ and $j$. $\partial{i}$ denotes neighbors of $i$ and $\delta_{\langle ij \rangle}$ denotes particles that are in contact. We ignore translational modes and rattlers, which trivially correspond to zero modes of the dynamical matrix.

To apply Lees-Edwards boundary conditions, we use periodicity vectors $\{\vec{L_x},\vec{L_y}\}$. After initializing the system, we keep the simulation area fixed at every strain $\gamma$ and we choose to strain along the $x$ direction, such that $\vec{L_x}(\gamma) = \vec{L_x}(0)$ and $\vec{L_y}(\gamma) = \vec{L_y}(0) + \gamma\vec{L_y}(0)\cdot\hat{x}$.

At each value of applied strain $\gamma$, we calculate the resulting stress tensor $\sigma_{\alpha \beta}$ via the Born-Huang approximation \cite{born_dynamical_1954}.

\begin{equation}
\sigma_{\alpha \beta} = -\frac{\phi}{N}\sum_{k}(r^{(k)}_\alpha f_\beta^{(k)}),
\label{eqn:bornHuang}
\end{equation}

\noindent
where the sum is over all bonds $k$ with distance vector $\vec{r}^{(k)}$ and force vector $\vec{f}^{(k)}$. The shear stress that we report is $\sigma_{xy}$, and the pressure is  $p = -\frac{1}{D}\sum_{\alpha}\sigma_{\alpha \alpha}$ \cite{ohern_jamming_2003}. \\

\noindent
{\em Numerical calculation of the shear modulus---} The shear modulus $G$ is \cite{maloney_amorphous_2006, merkel_geometrically_2018}:

\begin{equation}
G = \frac{1}{L^2}\bigg[ \frac{\partial^2U}{\partial{\gamma}^2} - \sum_{\lambda_{q}\neq 0} \frac{1}{\lambda_{q}} \sum_{k\alpha} (C_{k}^{q\alpha} \frac{\partial^2U}{\partial{\gamma}\partial{r_{k}^{\alpha}}})^2\bigg]
\end{equation}
\noindent
where $\lambda_{q}$ and $C^{q}$ are the eigenvalues and eigenvectors of the Hessian. From this equation one can see that stress drops associated with discontinuities in the shear modulus occur when an eigenvalue $\lambda_{q}$ goes to zero.

To obtain the partial derivatives of the energy with strain, it is useful to write the distance between particles explicitly in $x$ and $y$-components in terms of the boundary conditions:

\begin{equation}
\rho_{ij}^{x} = \sigma_{i} + \sigma_{j} - r_{j}^{x} + r_{i}^{x} - q_{xij}L_{xx} - q_{yij}[L_{yx}(0)+\gamma L]
\end{equation}

\begin{equation}
\rho_{ij}^{y} = \sigma_{i} + \sigma_{j} - r_{j}^{y} + r_{i}^{y} - q_{yij}L_{yy}
\end{equation}
\noindent
where $q_{\alpha ij}$ is 0 if the particles are in the same simulation box, but it is $+1$ if particle $j$ is in a periodic box to the right ($q_{xij}$) or above $i$ ($q_{yij}$), and it is $-1$ if particle $j$ is in a periodic box to the left or below $i$.

This makes computing the derivatives straightforward

\begin{equation}
\frac{\partial{U}}{\partial{\gamma}} = -L \sum_{i<j} q_{yij} \varepsilon_{ij}^{\xi-1}n_{ij}^{x},
\end{equation}

\begin{equation}
\frac{\partial^2U}{\partial\gamma^2} = L^2 \sum_{i<j}q_{yij}^2\bigg[(\xi-1)\varepsilon_{ik}^{\xi-2}(n^{x}_{ik})^2 + \frac{\varepsilon_{ik}^{\xi-1}}{\rho_{ik}}[(n^{x}_{ik})^2-1]\bigg],
\end{equation}

\begin{equation}
\frac{\partial^2U}{\partial{\gamma}\partial{r_{k}^{\alpha}}} = L \sum_{i\in \partial{k}} q_{yik}\bigg[(\xi-1)\varepsilon_{ik}^{\xi-2}n^{x}_{ik}n^{\alpha}_{ik} + \frac{\varepsilon_{ik}^{\xi-1}}{\rho_{ik}}[n^{x}_{ik}n^{\alpha}_{ik}-\delta^{x\alpha}]\bigg],
\end{equation}
\noindent
where $j\in\partial i/B$ indicates particles that contact across the top boundary, such that particle $i$ is above the boundary and particle $j$ is below the boundary.
\\

\noindent
\em{Fraction of contact changes that are rearrangements ---} \em
\noindent
We can convert numbers of events to a relative fraction of rearrangements $\eta$ by assuming that we traverse a total strain $\Delta \gamma$ and then counting the expected numbers of each event:

\begin{equation}
\eta = \frac{N_{r}}{N_{r}+N_{n}} = \frac{\frac{\Delta \gamma}{\Delta \gamma_{r}}}{\frac{\Delta \gamma}{\Delta \gamma_{r}} + \frac{\Delta \gamma}{\Delta \gamma_{n}}} = \frac{1}{1+\frac{\Delta \gamma_{r}}{\Delta \gamma_{n}}}.
\end{equation}\\

\noindent
\subsection{Deformations required to produce a contact change} \label{merlijnTable}

Previous work by some of us~\cite{van_deen_contact_2014,van_deen_contact_2016} has studied the deformations required to produce a contact change in an arbitrary contact potential with power $\xi$. Here we reproduce and extend those arguments and discuss how the results presented in the main text are consistent with previous results.

First we consider pure compression. From O'Hern et al~\cite{ohern_jamming_2003}, we know that for an arbitrary contact potential the excess coordination $\Delta z$ scales with the pressure $p$ as  $\Delta z \sim p^{1/\beta}$ in the thermodynamic limit, where \mbox{$\beta = 2(\xi - 1)$} or $N\Delta z \sim (N^\beta p)^{1/\beta}$. 

Following conventions established by van Deen~\cite{van_deen_contact_2014}, $\varepsilon_{mk}$ denotes the compressional strain required to make a contact and $\varepsilon_{bk}$ denotes the compressional strain required to break a contact. The contact change strain $\varepsilon_{cc}$ is then the minimum of the absolute values of the making and breaking strains.

In the $N^\beta \gg 1$ regime, making \mbox{(+)} or breaking \mbox{(-)} a contact changes $N\Delta z/2$ by one, which generates a change in pressure $N \Delta z / 2 \pm 1 \sim [N^\beta (p \pm \delta p)]^{1/\beta}$. Then $\pm \frac{\delta}{\delta p}[N^\beta (p \pm \delta p)]^{1/\beta}] \sim \pm Np^{(3-2\xi)/2(\xi-1)}$ describes the change in contacts given a change in pressure.  Inverting this expression gives  $\delta p \sim \pm p^{(2\xi - 3)/2(\xi-1)}/N$, which is the change in pressure required to make a single contact change. The compressional strain is the pressure change divided by the bulk modulus K: $\varepsilon_{cc} \sim \pm \delta p/K$, where the bulk modulus scales as $K \sim p^{(\xi -2)/(\xi-1)}$ \cite{ohern_jamming_2003}, yielding $\varepsilon_{cc} \sim \pm p^{1/2(\xi - 1)}/N = p^{1/\beta}/N$. 

In the $N^\beta \ll 1$ regime, the arguments of van Deen et al.~\cite{van_deen_contact_2014,van_deen_contact_2016}, do not depend on the interaction potential, and so they carry through unaltered. In summary:

\begin{center}
\begin{tabular}{c|c|c|l}
    $\varepsilon_{bk}$ & $\varepsilon_{mk}$ & $\varepsilon_{cc}$ & Regime\\
    \hline
    $-p$ & $1/N^2$ & $p$ & $N^\beta p \ll 1$ \\
    $-p^{1/\beta}/N$ & $p^{1/\beta}/N$ & $p^{1/\beta}/N$ & $N^\beta p \gg 1$
\end{tabular}
\end{center}

\noindent
To reconstruct the shear argument, we note that \mbox{$G \sim p^{(\xi - 3/2)/(\xi-1)}$} for $N^\beta p \gg 1$ and $G \sim 1/N$ for $N^\beta \ll 1$,~\cite{ohern_jamming_2003,goodrich_finite-size_2012}. We also follow van Deen's argument that $\sigma_{cc} \sim P/N$ because the pressure sets the force scale in the system, and the stress should scale appropriately with system size. We assume linear response, giving $\gamma_{cc} \sim \sigma_{cc}/G$, which means:

\[ \gamma_{cc} \sim
\begin{cases}
\frac{p^{1/\beta}}{N} & N^\beta p \gg  1,\\
p & N^\beta p \ll  1. \\
\end{cases}
\]

\begin{figure}
\includegraphics[width=0.9\columnwidth]{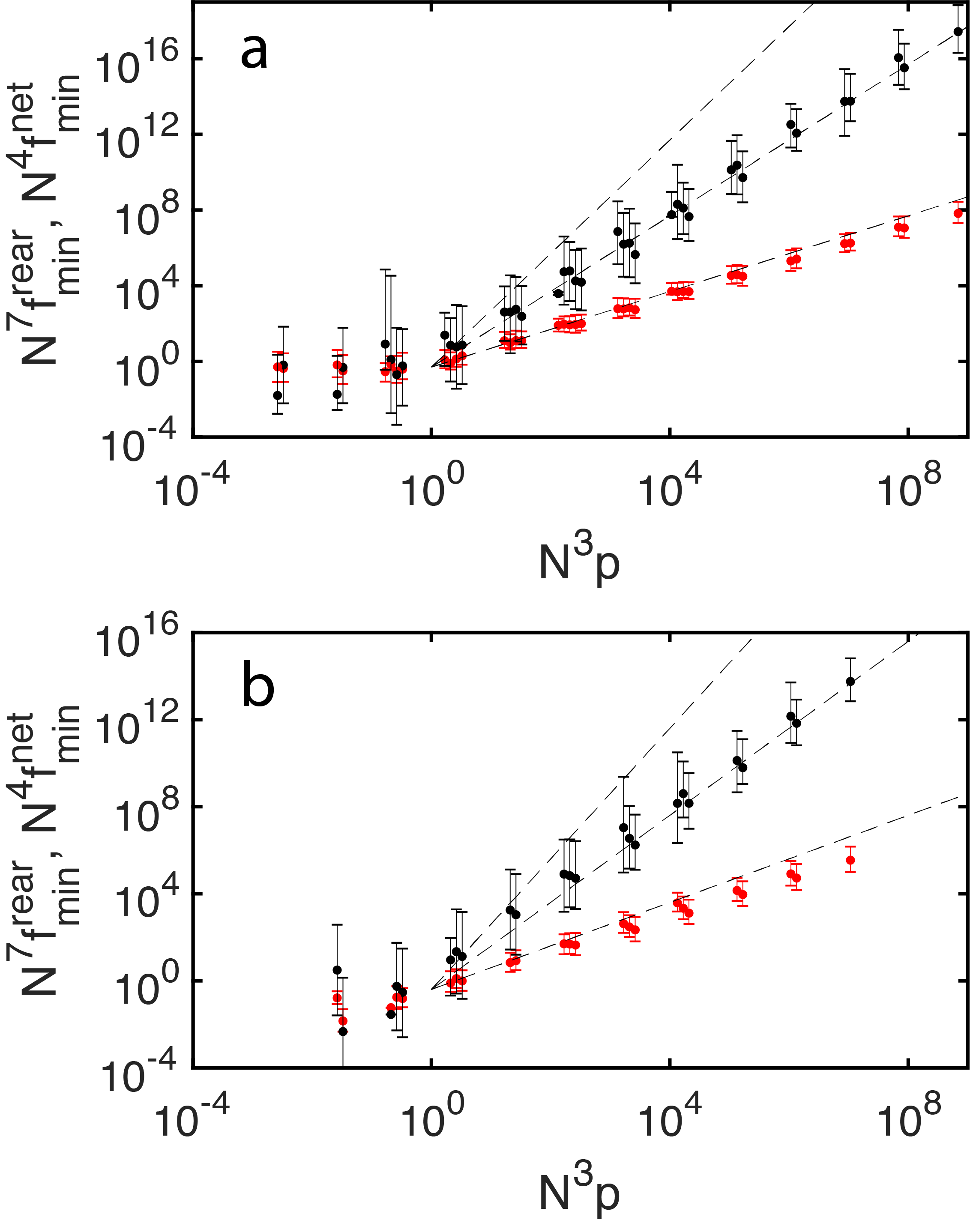}
\caption{The difference between $f_{min}$ for contact making network events (red) and rearrangements (black) for \textbf{a)} 2D Hertzian disks and \textbf{b)} 3D Hertzian spheres. In the $N^3p > 1$ regime, the scaling reduces to $f_{min}^{net} \sim p/N$ and $f_{min}^{rear} \sim p^2/N$.}
\label{fig:allHistSurfPlot}
\end{figure}

We note that the $N^\beta \gg 1$ regime predicted by van Deen agrees well with what we see, while the small system size prediction does not. However, a major differnce between the results of van Deen at al. and those presented here is that their results are shear stabilized, meaning the energy is minimized with respect to box size in addition to particle positions, and ours are not.  Perhaps as a consequence, van Deen et al. find that the first contact change is quite a bit more likely to be a contact breaking event, while ours are more often contact making (Fig. \ref{fig:allHistSurfPlot}). We expect this difference would cause significantly different behavior in the $N^\beta \ll 1$ regime, as observed.

\subsection{Supplemental Figures}
\FloatBarrier

\begin{figure}
\includegraphics[width=0.9\columnwidth]{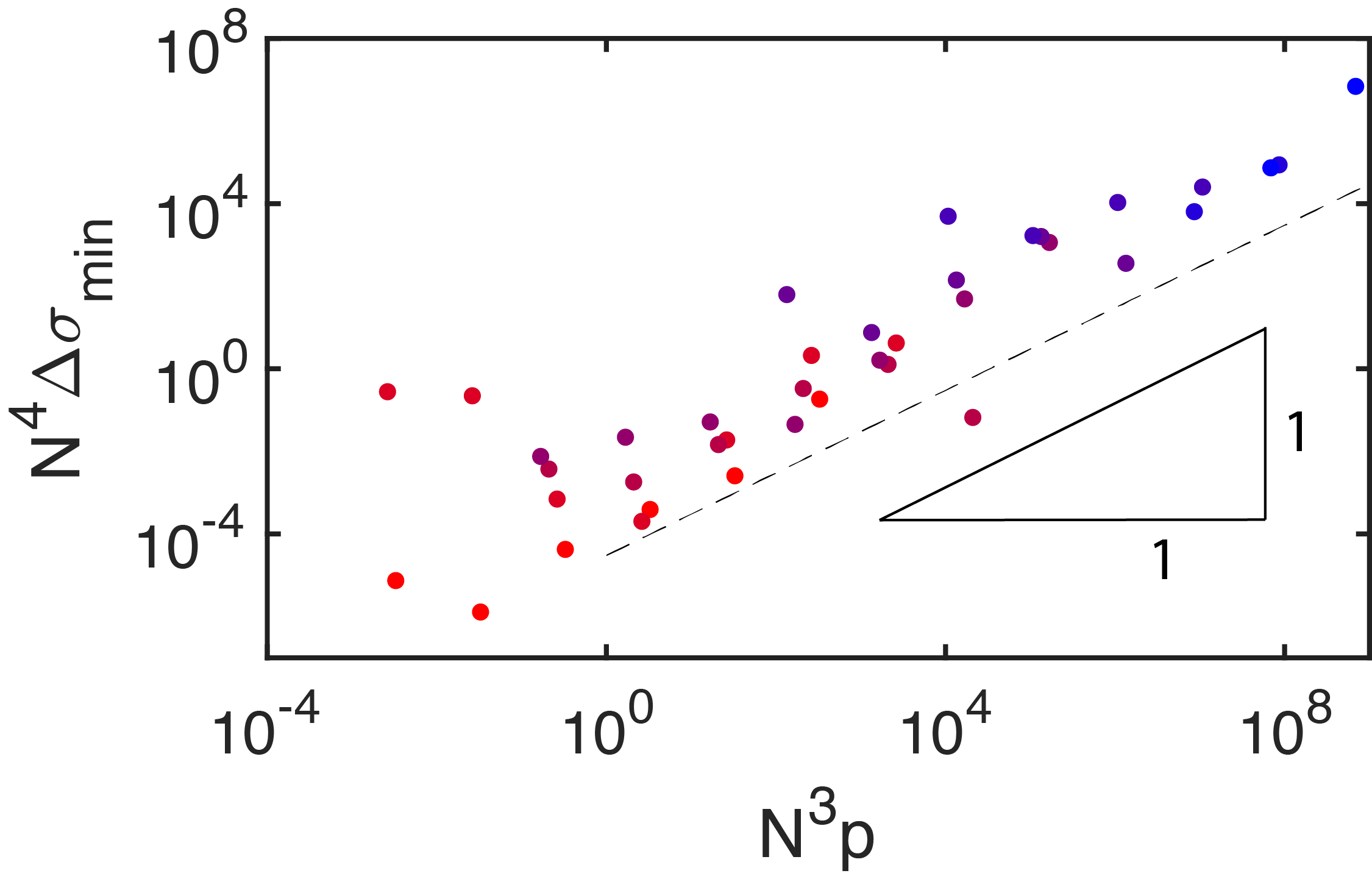}
\caption{The minimum observed stress drop associated with rearrangements as a function of $N$ and $p$ for 2D Hertzian systems. This is well defined at our level of precision for all values of $N$ and $p$, because the stress drops are always bimodally distributed. Because each is taken as a minimum observation within a distribution, it overestimates the lower bound on stress drops.}
\label{fig:stressDropLowerBound}
\end{figure}

\begin{figure*}
\includegraphics[width=0.9\textwidth]{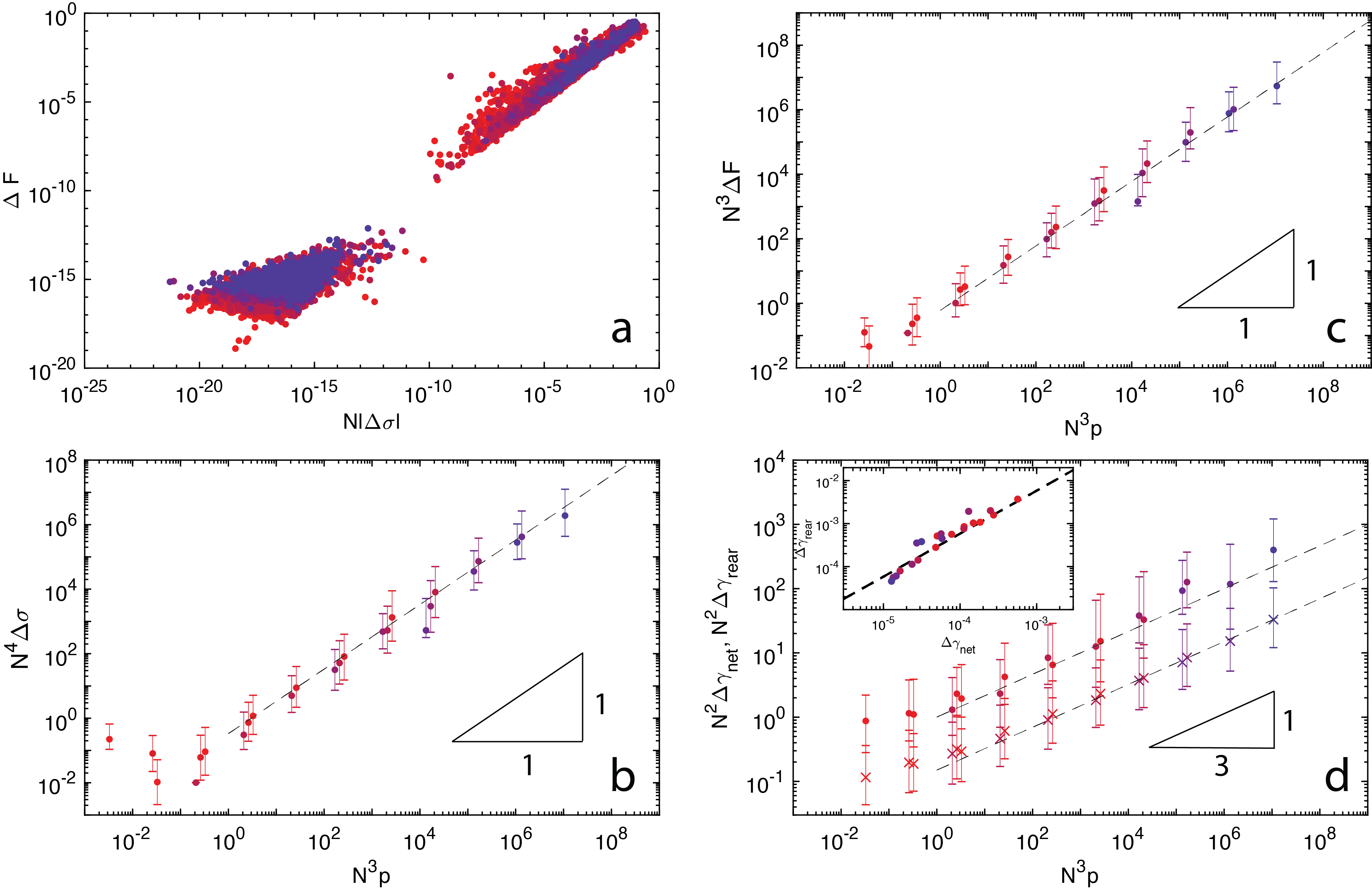}
\caption{Results for 3D Hertzian spheres. (\textbf{a}) A metric of reversibility versus the absolute value of the stress drop. We see a clear distinction between network events and rearrangements: network events lie in the lower left quadrant, are perfectly reversible, and have zero stress drop, while rearrangements lie in the upper right quadrant, are irreversible, and have a nonzero stress drop. The average stress drop (\textbf{b}) and the squared metric of reversibility (\textbf{c}) associated with rearrangements for each system size and pressure collapses onto a master curve when we plot these versus $N^\beta p$. Dashed lines represent the power law for each, and are shown as a guide to the eye. (\textbf{d}) Strain steps between rearrangements (circles) and network events (crosses). (Inset) A simple regression confirms that these two curves have the same slope, which implies a constant fraction of rearrangements across system size and pressure. A dashed line with the best fit from linear regression is plotted. Error bars on all plots are given as the middle 60\% of each distribution, and the colors are N = 32(red), 64, 128, 256, 512, and 1024(blue) with an even gradient.}
\label{fig:crosshair3d}
\end{figure*}

\begin{figure}
\includegraphics[width=\columnwidth]{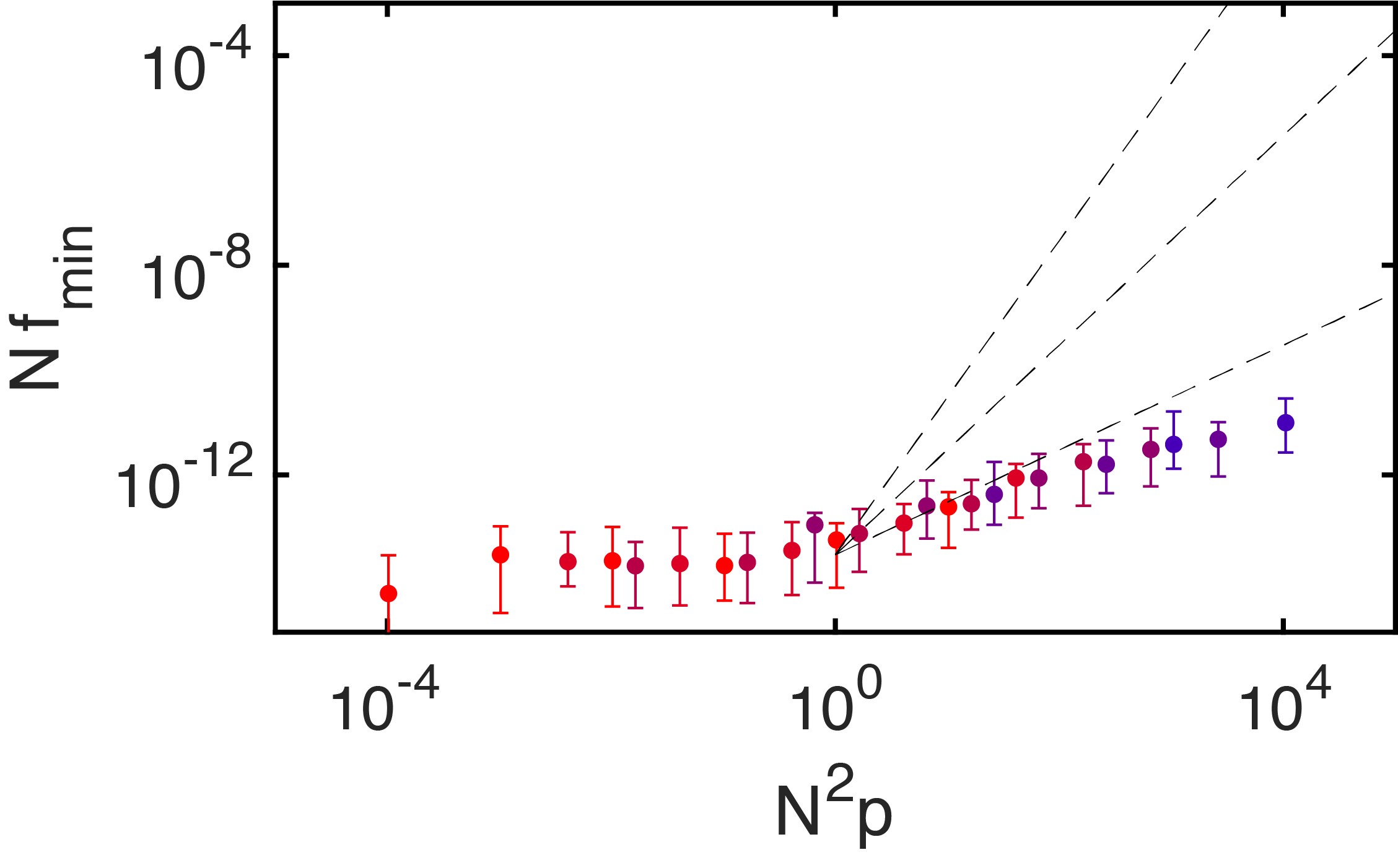}

\caption{The minimum force before a rearrangement in 2D Hookean disks. Dashed lines represent a slope of 1, 2, and 3. Our data clearly does not have any of these scalings, and is consistent with a force being below the numerical threshold for zero. Error bars represent the middle 60\% of each distribution, and the point represents the geometric mean. The colors vary with system size as N = 32(red), 64, 128, 256, 512, 1024, 2048, and 4096 (blue) with an even gradient.}
\label{fig:minForce2d}
\end{figure}

\bibliography{supplement}